\magnification=\magstephalf
%input macro
%at the start of a tex file. Note that different printers may or
%may not automatically include an offset---if the text is off to
%one side adjust the commands \hoffset and \voffset by adding or
%removing a comment sign %.
\def\etal{{et al.\ }}
\def\chaphead{}
\def\ni{\noindent}

\font\tfont=cmbxti10
\font\tenrm=cmr10
\font\tenit=cmti10
\font\eightmit=cmmi8
\font\eightrm=cmr8
\font\tenmit=cmmi10
\def\absmath{\textfont0=\tenrm \scriptfont0=\eightrm
	      \textfont1=\tenmit \scriptfont1=\eightmit}
\def\absfont{\let\rm=\tenrm \let\it=\tenit \rm\absmath}
\font\twelverm=cmr12
\font\twelveit=cmti12
\font\tenrm=cmr10
\font\twelvemit=cmmi12
\font\tenmit=cmmi10
\def\regmath{\textfont0=\twelverm \scriptfont0=\tenrm
	      \textfont1=\twelvemit \scriptfont1=\tenmit}
\def\peterfont{\let\rm=\twelverm \let\it=\twelveit \rm\regmath}
\def\pagenumbers{\headline={\hss --~\folio~--\hss}\footline{}\vskip 12pt} 
%restarts numbering after \nopagenumbers

%following macro defines vectors using arrows
  %This contradicts tex82 which has \b meaning
%underline instead of vector superscript. Skew is added to
%improve location of arrow.
%following macro defines vectors as boldface italic. Note in this case that
%all characters regarded as being in the argument of the macro will be
%boldface.
\newfam\vecfam

\textfont\vecfam=\tfont \scriptfont\vecfam=\seveni
\scriptscriptfont\vecfam=\fivei

 %error function

\def\spose#1{\hbox to 0pt{#1\hss}}

\font\tenrm=cmr10

\def\s{\ifmmode \widetilde \else \~\fi} %produces tilde in mathmode or
%horizontal mode.
     
%\def\={\overline}
\def\section{\S}
\newcount\notenumber
\notenumber=1
\newcount\eqnumber
\eqnumber=1
\newcount\fignumber
\fignumber=1
\newbox\abstr

%\numberpara produces numbered paragraphs with extra space and no indentation

\def\kmsmpc{{\rm\,km\,s^{-1}\,Mpc^{-1}}}
\def\msun{{\rm\,M_\odot}}

\def\c2m{{\rm\,cm^{-2}}}
\def\3cm{{\rm\,cm^{-3}}}
\def\gcm3{{\rm\,g\,cm^{-3}}}
\def\s{{\rm\,s}}

\def\funits{{\rm\,ergs\,cm^{-2}\,s^{-1}\,Hz^{-1}}} 
\def\uvunits{{\rm\,ergs\,cm^{-2}\,s^{-1}\,Hz^{-1}\,sr^{-1}}}

%\note macro produces sequentially numbered footnotes at bottom of page
%\foot macro produces sequentially numbered footnotes inserted in text
\def\note#1{\footnote{$^{\the\notenumber}$}{#1}\global\advance\notenumber by 1}
\def\foot#1{\raise3pt\hbox{\tenrm \the\notenumber}
     \hfil\par\vskip3pt\hrule\vskip6pt
     \noindent\raise3pt\hbox{\tenrm \the\notenumber}
     #1\par\vskip6pt\hrule\vskip3pt\noindent\global\advance\notenumber by 1}

%\abstract macro makes abstracts
\def\abstract#1{\setbox\abstr=\vbox{\hsize 5.0truein{\par\noindent#1}}
    \centerline{ABSTRACT} \vskip12pt \hbox to \hsize{\hfill\box\abstr\hfill}}
     
%\Dt and \dt put Newton's notation dots above upper and lower case chars
\def\Dt{\spose{\raise 1.5ex\hbox{\hskip3pt$\mathchar"201$}}}    % upper case
\def\dt{\spose{\raise 1.0ex\hbox{\hskip2pt$\mathchar"201$}}}    % lower case

% equation numbering
%\new macro produces sequentially numbered equations by writing \eqno(\new)
%at end of displayed equations
\def\new{{\rm\chaphead\the\eqnumber}\global\advance\eqnumber by 1}
\def\last{\advance\eqnumber by -1 {\rm\chaphead\the\eqnumber}\advance
     \eqnumber by 1}
%to name an equation, place command "\eqnam{\Poisson}" before equation, and
%thereafter "equation(\Poisson)" will generate the proper equation number.
\def\eqnam#1{\xdef#1{\chaphead\the\eqnumber}}
     
%figure numbering
%\nfig macro assigns number to a figure
\def\nfig{\chaphead\the\fignumber\global\advance\fignumber by 1}
%\nfiga permits a,b,c etc. to be added to figure number
\def\nfiga#1{\chaphead\the\fignumber{#1}\global\advance\fignumber by 1}
\def\rfig#1{\advance\fignumber by -#1 \chaphead\the\fignumber
     \advance\fignumber by #1}
\def\fignam#1{\xdef#1{\chaphead\the\fignumber}}
%reference macros. To generate reference to a paper in Ap.J. volume 300, p.123
%write \apj{Claus, S. 1990,}{300}{123}
\def\refindent{\par\noindent\parskip=4pt\hangindent=3pc\hangafter=1 }

\def\apjj#1#2#3{\refindent#1,  {ApJ}, {#2}, #3}

\def\mnrass#1#2#3{\refindent#1,  {MNRAS}, {#2}, #3}

\def\ajj#1#2#3{\refindent#1,  {AJ}, {#2}, #3}

\def\aaa#1#2#3{\refindent#1,  {A\&A}, {#2}, #3}

\def\sectionbegin#1{\vskip6pt\par\noindent{\centerline{\bf#1}}\par\vskip4pt}
\def\subsectionbegin#1{\vskip5pt\par\noindent{\centerline{\it#1}}\par
\vskip3pt}
\def\subsubsectionbegin#1{\vskip4pt\par\noindent{\centerline{\it#1}}\par
\vskip2pt}

%\lta and \gta produce > and < signs with twiddle underneath
\def\lta{\mathrel{\spose{\lower 3pt\hbox{$\mathchar"218$}}
     \raise 2.0pt\hbox{$\mathchar"13C$}}}
\def\gta{\mathrel{\spose{\lower 3pt\hbox{$\mathchar"218$}}
     \raise 2.0pt\hbox{$\mathchar"13E$}}}
     
%\sec produces arcsec symbol so that 3\sec5 produces 3."5 with the second
%symbol and the period aligned.

%\magnification=\magstephalf
%\magnification=\magstep1
\hoffset=-0.3truein
%\voffset=0.8truein
\parskip=2pt

\hsize=7.1truein 
\vsize=9.0truein
\overfullrule=0pt	% delete the nasty little black boxes for overfull box
\def\ref#1{Ref.~#1}			% 	for inline references

\def\HI{\hbox{H~$\scriptstyle\rm I\ $}}

\def\HeI{\hbox{He~$\scriptstyle\rm I\ $}}
\def\HeII{\hbox{He~$\scriptstyle\rm II\ $}}

\def\CIV{\hbox{C~$\scriptstyle\rm IV\ $}}

\def\nHI{_{\rm HI}}

\def\cmm{\,{\rm cm}^{-2}}

\def\Ha{H$\alpha\ $}
\def\Lya{Ly$\alpha\ $}
\def\Lyb{Ly$\beta\ $}
\def\msunits{{\rm\,M_\odot\,yr^{-1}\,Mpc^{-3}}}
\def\munits{{\rm\,M_\odot\,Mpc^{-3}}}
\def\ldunits{{\rm\,ergs\,s^{-1}\,Mpc^{-3}}}

\def\sunits{{\rm\,ergs\,s^{-1}\,Hz^{-1}\,Mpc^{-3}}}

\def\page{\vfill\eject}
\def\ub{U_{300}-B_{450}}
\def\bv{B_{450}-V_{606}}
\def\vi{V_{606}-I_{814}}
\def\bi{B_{450}-I_{814}}
\overfullrule=0pt
\nopagenumbers

\centerline{\bf HIGH REDSHIFT GALAXIES IN THE HUBBLE DEEP FIELD. COLOR}
\centerline{{\bf SELECTION AND STAR FORMATION HISTORY TO ${\bf z\sim 4}$}
\note{Based on observations with the NASA/ESA {\it Hubble Space Telescope}
obtained at the Space Telescope Science Institute which is operated by AURA
under NASA contract NAS 5-2655.}}
\bigskip
\bigskip
\centerline{{\it Piero Madau, Henry C. Ferguson, Mark E. Dickinson}}
\centerline{Space Telescope Science Institute, 3700 San Martin Drive,
Baltimore MD 21218, USA}
\centerline{e-mail: madau@stsci.edu, ferguson@stsci.edu, med@stsci.edu}
\medskip
\centerline{{\it Mauro Giavalisco}\note{Hubble Fellow}}
\centerline{Carnegie Observatories, 813 Santa Barbara Street, Pasadena, Ca 
91101-1292}
\centerline{e-mail: mauro@ociw.edu}
\medskip
\centerline{{\it Charles C. Steidel}\note{Alfred P. Sloan Foundation Fellow}
$^,$\note{NSF Young Investigator}}
\centerline{Palomar Observatory, California Institute of Technology, 
Mail Stop 105-24, Pasadena, CA 91125}
\centerline{e-mail: ccs@astro.caltech.edu}
\medskip
\centerline{{\it Andrew Fruchter}}
\centerline{Space Telescope Science Institute, 3700 San Martin Drive,
Baltimore MD 21218}
\centerline{e-mail: fruchter@stsci.edu}

\bigskip
\centerline{\bf ABSTRACT}

The Lyman decrement associated with the cumulative effect of \HI in QSO
absorption systems along the line of sight provides a distinctive feature
for identifying galaxies at $z\gta2.5$. Color criteria, which are sensitive to
the presence of a Lyman-continuum break superposed on an otherwise flat UV
spectrum, have been shown, through Keck spectroscopy, to successfully identify 
a substantial population of star-forming galaxies at $3\lta z\lta 3.5$
(Steidel \etal 1996a). Such
objects have proven surprisingly elusive in field-galaxy redshift surveys;
quantifying their surface density and morphology is crucial for determining how
and when galaxies formed. The {\it Hubble Deep Field} (HDF) observations offer
the opportunity to exploit the ubiquitous effect of intergalactic absorption
and obtain useful statistical constraints on the redshift distribution of
galaxies considerably fainter than current spectroscopic limits. We model the
\HI cosmic opacity as a function of redshift, including scattering in 
resonant lines of the Lyman series and Lyman-continuum absorption, and use
stellar population synthesis models with a wide variety of ages, metallicities,
dust contents, and redshifts, to derive color selection criteria that provide a
robust separation between high redshift and low redshift galaxies. From the HDF
images we construct a sample of star-forming galaxies at $2\lta z\lta 4.5$.
While none of the $\sim 60$ objects in the HDF having known Keck/LRIS 
spectroscopic redshifts in the range $0\lta z\lta 1.4$ is found to contaminate
our high-redshift sample, our color criteria are able to efficiently select the
$2.6\lta z\lta 3.2$ galaxies identified by Steidel \etal (1996b). 

The ultraviolet (and blue) dropout technique opens up the possibility of
investigating cosmic star and element formation in the early universe. We set a
lower-limit to the ejection rate of heavy elements per unit comoving volume
from Type II supernovae at $\langle z\rangle=2.75$ of $\approx 3.6\times
10^{-4}\msunits$ (for $q_0=0.5$ and $H_0=50\kmsmpc$), which is 3 times
higher than the local value, but still 4 times lower than the rate observed at
$z\approx 1$. At $\langle z\rangle=4$, our lower limit to the cosmic metal
ejection rate is $\approx 3$ times lower than the $\langle z\rangle=2.75$
value. We discuss the implications of these results on models of galaxy
formation, and on the chemical enrichment and ionization history of the
intergalactic medium. 

\medskip
\ni{\it Subject~headings}: cosmology: observations -- galaxies: evolution -- 
intergalactic medium -- quasars: absorption lines -- ultraviolet: galaxies

\page
\pagenumbers 

\sectionbegin{1. INTRODUCTION}

Much observing time has been devoted in the past few years to the problem of
the detection of galaxies at high redshift,  as it is anticipated that any
knowledge of their early luminosity and color evolution will put constraints on
the history of structure and metal formation. While it has become clear that
blank-sky surveys for strong \Lya-emitting primeval galaxies are not
particularly efficient (e.g., Pritchet \& Hartwick 1990; Djorgovski \&
Thompson 1992), the method of obtaining multicolor broadband observations  of
the emitter's rest-frame UV stellar continuum has been successfully applied to
detect galaxies at cosmological distances (Steidel \etal 1996a).

Ground-based observations have used  color techniques which are sensitive to
the presence of a Lyman-continuum break to identify or set limits on the number
of high redshift galaxies (Steidel \& Hamilton 1992, 1993). The signature of a
distant, star-forming galaxy in such surveys is a very red $U-B$ color,
combined with colors in longer-wavelength filters that are much bluer. The
early work by Guhathakurta \etal (1990) showed that Lyman-break objects do not
dominate the galaxy counts at faint apparent magnitudes. Assuming that young,
star-forming galaxies have flat spectra longward of the Lyman-break, Steidel \&
Hamilton (1992, 1993), and Steidel, Pettini, \& Hamilton (1995) have designed
and used a custom set of broad-band filters to select high-$z$ galaxies in the
fields of distant QSOs. Recent deep spectroscopy with the W. M. Keck telescope
has shown the high efficiency of such color-selection technique: Steidel \etal
(1996a) have identified 17 Lyman-break star-forming galaxies with redshifts
$3.0\lta z \lta 3.5$, which are most likely the progenitors of the present-day
bright spirals and ellipticals observed during the assembly of their cores
(Giavalisco, Steidel, \& Macchetto 1996). 

With its high spatial resolution, the {\it Hubble Space Telescope} (HST) offers
the opportunity to study the faint galaxy population in unprecedented detail.
In particular, the images of the {\it Hubble Deep Field} (HDF), obtained with
the Wide Field Planetary Camera (WFPC-2) during December 1995, represent the
deepest optical imaging survey undertaken so far (Williams \etal 1996),
reaching 5-$\sigma$ limiting AB magnitudes of roughly 27.7, 28.6, 29.0, and
28.4 (for an aperture area of 0.2 square arcsec) in the F300W, F450W, F606W,
and F814W bandpasses (the number corresponds to the central wavelength in nm),
respectively. In this paper, we describe selection criteria for this filter
system that can be used to identify a sample of likely precursors to present
day galaxies at $z>2$, with little contamination from low redshift galaxies.
Our strategy is designed to exploit the effect of the increasing opacity of the
intergalactic medium (IGM) at high redshifts, as indicated by the plethora of
absorption lines seen in the spectra of background quasars: enough neutral
hydrogen is known to be contained in the clumps of highly ionized gas which
form the \Lya forest, and in the metal-line absorption systems associated with
the halo regions of intervening bright galaxies, to significantly attenuate the
UV flux from distant sources. We show that the cumulative effect of \HI in QSO
absorption systems along the line of sight provides a distinctive feature for
identifying star-forming galaxies in the HDF.  Although other spectral
features, such as the 4000 and $912\,$\AA\ breaks which characterize the
integrated spectra of stellar populations, with the latter possibly enhanced by
self-absorption from interstellar gas within the galaxy itself, may also be
used as tracers of redshifts, the model predictions of their magnitude are
sensitive to the unknown physical and evolutionary state of the galaxy, i.e.,
its star formation history, age, and \HI distribution, and hence are subject to
substantial uncertainties.  By contrast, the ``reddening'' effect due to atomic
processes in cosmological distributed QSO absorption systems is ubiquitous,
quite strong for $z\gta2.5$, and can be reliably taken into account. Although
stochastic in nature, r.m.s. fluctuations away from the mean opacity are bound
to be modest in most situations, due to the broadband nature of the adopted
filter set. 

The technique to analyze the HDF images we develop in this paper is based on the
theoretical study by Madau (1995) (see also Yoshii \& Peterson 1994), and is an
extension of the Lyman-break color criterion developed by Steidel \& Hamilton 
(1992, 1993), and Steidel, Pettini, \& Hamilton (1995). We extend this technique
to the HDF filter system, and perform several experiments with simulated and
real galaxy spectra to define effective selection criteria for high redshift
galaxies, and identify possible sources of contamination from low redshifts. By
computing colors for an extremely wide range of model galaxy spectra, we are
able to tune the criteria to provide what we believe are largely uncontaminated
samples of star-forming galaxies in the redshift ranges $2<z<3.5$ and
$3.5<z<4.5$, and use these samples to estimate the integrated star formation
and metal ejection rates at those redshifts. We show how follow-up spectroscopy
with Keck/LRIS by Steidel \etal (1996b), the Hawaii$+$Caltech group (Cohen
\etal 1996), and the Berkeley HDF group (Moustakas, Zepf, \& Davis 1996)
supports the efficiency of our color selection techniques. Future work on high
redshift galaxies in the HDF will investigate the luminosity function, study
the morphologies, and compare the numbers and properties of high-$z$ candidates
to the predictions of a variety of galaxy-evolution models. 

Throughout this paper, unless otherwise stated we shall adopt a flat cosmology
with $q_0=0.5$ and $H_0=50\kmsmpc$. 

\sectionbegin{2. INTERGALACTIC ATTENUATION}

In this section we will briefly review the theory of the propagation of UV
radiation through a clumpy universe, following Madau (1995).

\subsectionbegin{2.1. Basic Equations}

Let $L(\nu_{\rm em})$ be the specific power emitted with frequency $\nu_{\rm 
em}$ by a source at redshift $z_{\rm em}$. The mean specific flux observed at 
Earth is
$$
\langle f(\nu_{\rm obs})\rangle ={(1+z_{\rm em})L(\nu_{\rm em})\over 4\pi
d_L^2} \langle e^{-\tau}\rangle, \eqno(\new)
$$
where $\nu_{\rm obs}=\nu_{\rm em}/(1+z_{\rm em})$, $d_L$ is the luminosity
distance to $z_{\rm em}$, and the average transmission over all lines of sight
is, assuming Poisson-distributed clouds, \eqnam{\transm} 
$$
\langle e^{-\tau}\rangle=\exp\left\{\int_0^{z_{\rm em}}\int 
{{\partial^2N}\over {\partial N\nHI\partial
z}}\left[1-e^{-\tau_c}\right] dN\nHI dz\right\}. \eqno(\new)
$$
Here, $\tau_c$ is the optical depth through an individual cloud at frequency
$\nu=\nu_{\rm obs}(1+z)$, and $(\partial^2N/ \partial N\nHI\partial z)$ is
the redshift and column density distribution of absorbers along the path. An
\lq\lq effective'' optical depth of a clumpy medium can be defined as
$\tau_{\rm eff}=-\ln(\langle e^{-\tau}\rangle)$. 

Along with resonant line scattering from Ly$\alpha$, $\beta$,
$\gamma$, and higher order members, we include photoelectric absorption from
\HI in the \Lya forest clouds and Lyman-limit systems along the line
of sight. Since the bluest filter used for the HDF observations is centered 
at $3000\,$\AA\ and has FWHM$\sim 800\,$\AA, galaxies will only be subject to
\HI cosmological attenuation from material located at $z\gta (2600/1216)-1=
1.1$. 

\subsectionbegin{2.2. Line Blanketing {\rm \&} Continuum Absorption}

For $\lambda_{\beta}(1+z_{\rm em}) < \lambda_{\rm obs} < \lambda_{\alpha} 
(1+z_{\rm em})$, where $\lambda_{\alpha}=1216\,$\AA\ and
$\lambda_{\beta}=1026\,$\AA, a galaxy's continuum intensity is attenuated by
the combined blanketing effect of many \Lya forest absorption lines, with
effective opacity 
$$ 
\tau_{\rm eff}=0.0036\left({\lambda_{\rm obs}\over \lambda_{\alpha}}\right)
^{3.46}, \eqno(\new) 
$$ 
(Press, Rybicki, \& Schneider 1993). Hence, line blanketing from \Lya alone will
produce $\gta 1\,$mag of attenuation in the continuum spectrum shortward of 
$6000\,$\AA\ of a galaxy at $z_{\rm em}\gta 4$.

When $\lambda_{\rm obs}\le\lambda_{\beta}(1+z_{\rm em})$, a significant
contribution to the blanketing opacity comes from the higher order lines of the
Lyman series. A standard curve of growth analysis has been applied to
numerically compute the attenuation expected from line blanketing of Ly$\beta$,
$\gamma$, $\delta$ plus 13 higher order members. In the wavelength
range $\lambda_{i+1}(1+z_{\rm em})<\lambda_{\rm obs}<\lambda_i(1+z_{\rm em})$,
the total optical depth can be written as the sum of the contributions from the
$j\rightarrow 1$ transitions, 
$$
\tau_{\rm eff}=\sum_{j=2,i} A_j({\lambda_{\rm obs}\over
\lambda_j})^{3.46}, \eqno(\new)
$$ 
where $A_j=(1.7\times 10^{-3}, 1.2\times 10^{-3},9.3\times 10^{-4})$, and
$\lambda_j=(1026, 973$, $950\,$\AA) for Ly$\beta$, $\gamma$, and $\delta$, 
respectively.  It is worth noting that heavy element absorbers make a
negligible contribution to the blanketing optical depth at high redshifts, 
as this is dominated by those lines which lie at the transition between 
the linear and the flat part of the curve of growth, i.e., with $N\nHI
\approx 10^{13.6}\cmm$ in the case of \Lya.

Continuum absorption by \HI affects photons observed at
$\lambda_{\rm obs}<\lambda_L(1+z_{\rm em})$, where $\lambda_L=912\,$\AA\ is the
Lyman limit. An approximate (within 5\%) integration of equation (\transm)
yields for the effective photoelectric optical depth along the line of sight:
$$
\tau_{eff}=0.25x_c^3 (x_{em}^{0.46}-x_c^{0.46})+9.4x_c^{1.5}
(x_{em}^{0.18}-x_c^{0.18})-0.7x_c^3(x_c^{-1.32}-x_{em}^{-1.32})-0.023
(x_{em}^{1.68}-x_c^{1.68}), \eqno(\new)
$$
where $x_c=(\lambda_{\rm obs}/\lambda_L)$ for $\lambda_{\rm obs}>\lambda_L$,
and $x_{em}=1+z_{em}$. The first term on the right-hand side represents the
approximate contribution of Lyman-$\alpha$ clouds, the others are due to 
Lyman-limit systems. Absorbers with $N\nHI\sim 10^{17}\cmm$ dominate the 
cosmic continuum opacity.\note{The \HeI contribution to the attenuation is
negligible in the case of a QSO-dominated ionizing background, while \HeII
absorption on the way must be included if $(1+z_{\rm em})>\lambda_{\rm
obs}/228$\AA\ (e.g., Madau 1992).} 

While our formula for the line blanketing optical depth is entirely
consistent with the flux deficits $D_A$ and $D_B$ observed in the spectra of
QSOs below their \Lya and \Lyb emission features (e.g., Schneider,
Schmidt, \& Gunn 1991), the derived continuum opacity is subject to significant
uncertainties, as very limited information exist on optically thin absorbers
with $10^{16}\lta N\nHI\lta 10^{17}\cmm$. 

\subsectionbegin{2.3. Cosmic Transmission}

Figure 1{\it a} shows the characteristic staircase profile (cf M{\o}ller
\& Jakobsen 1990) of the mean cosmic transmission $\langle e^{-\tau}\rangle$
for a source at $z_{\rm em}=3$, 4, and 5, as a function of observed wavelength.
Also plotted are the response functions of the photometric system we shall
adopt in our discussion. This consists of the four broad {\it HST} passbands
F300W, F450W, F606W, and F814W (roughly $UBVI$). These filters efficiently
cover most of the bandpass accessible to the {\it HST} WFPC-2. 

It is clear that the observed broadband colors of cosmological distant objects
will be strongly \lq\lq reddened" by continuum absorption and blanketing of
discrete absorption lines which move into and out of the color passbands with
changing $z$. To be quantitative, we must take into account that the mean
transmission observed is not $\langle e^{-\tau}\rangle$ but rather an average
over the bandpass, 
$$ 
Q(z_{\rm em})=\int e^{-\tau_{\rm eff}}T(\lambda)d\lambda,  \eqno(\new)
$$ 
where $T(\lambda)$ is the normalized transmittance of the relevant filter. In
the following, we shall measure the mean integrated intergalactic attenuation
in magnitudes, $\Delta m=-1.086\ln Q$. Figure 1{\it b} shows $\Delta U_{300}$,
$\Delta B_{450}$, $\Delta V_{606}$, and $\Delta I_{814}$, 
the observed magnitude increases at the corresponding bandpass due to
intervening absorption, as a function of emission redshift in the redshift
range $1.5<z<5$. [These increments must be added to the term $-1.086\ln(1+z_{\rm
em})$ to get the standard $K$-correction for a flat emitted spectrum.] 

We are now able to assess how intergalactic absorption will modify the intrinsic
photometric properties of galaxies at high redshift. At $z_{\rm em}\approx 1.5$,
\Lya line blanketing starts to cause a small apparent depression in the 
$U_{300}$-band continuum. As higher order lines of the Lyman series move into
the bandpass, $\Delta U_{300}$ increases rapidly until, at $z_{\rm em}=2.3$, 
sources in the background appear, on average, $\approx 0.7\,$ mag fainter
in the ultraviolet. By $z_{\rm em}=3$, Lyman-continuum absorption from
forest clouds and Lyman-limit systems is so severe that $\Delta U_{300}\approx
2.8\,$mag (of which $\approx 1.9\,$mag are due to \Lya clouds alone): 
galaxies will either appear very red in $U_{300}-B_{450}$ or will drop out of
the F300W image altogether.\note{Note that $\Delta U_{300}$ flattens for $z_{\rm
em}\gta 3.6$ because of a significant ``red-leak'' of the F300W filter above
$6700\,$\AA.}~ Similarly, the \Lya line progressively enters the F450W band
beyond $z_{\rm em}\approx 2.7$. At $z_{\rm em}=3.8$, line blanketing from
the Lyman series produces $\Delta B_{450}\approx 1.2\,$mag. By $z_{\rm
em}=4.4$, Lyman-continuum absorption contributes significantly to the
total opacity, and $\Delta B_{450}$ exceeds $3.3\,$ mag. The F606W bandpass is
only weakly affected by intergalactic absorption for $z_{\rm em}\lta 4$. Hence,
again, star-forming galaxies at $z_{\rm em}\sim 4$ will either appear very red
in $B_{450}-V_{606}$ or will effectively be undetectable in the blue band. By
$z_{\rm em}=4.6$, line blanketing from the Lyman series produces $\Delta
V_{606}\approx 1\,$mag. As \HI absorption does not affect the F814W passband
for $z_{\rm em}\lta 5.7$, $\Delta I_{814}=0$ in the redshift range of
interest here. 

Although Figure 1{\it b} refers to the average accumulated absorption only,
r.m.s. fluctuations away from the mean transmission are predicted to be
relatively small after integration over broad bandpasses (Press \etal 1993).
For example, the observed $B_{450}$-band continuum of a galaxy at $z_{\rm
em}\gta 3$ is attenuated by the blanketing of more than 90 Ly$\alpha$ lines 
with $W>0.3$\AA\ along the light travel path. At these redshifts, excursions
away from the average curve will be the largest in the F300W filter, due to the
contribution to the photoelectric opacity by the rarer, optically thick
Lyman-limit systems. Even in this case, however, with $dN/dz\approx
0.27(1+z)^{1.55}$ absorbers per unit redshift with $N\nHI>1.6\times
10^{17}\cmm$ (Storrie-Lombardi \etal 1994), as many as $\Delta N\sim 1.5$
Lyman-limit systems along a random path to $z_{\rm em}\gta 2.7$ would be
expected, on average, to photoelectrically absorb photons in the ultraviolet
bandpass. As the probability of intersecting at least one metal system is
large, $1-e^{-\Delta N}\approx 80$\%, we expect only one out of five
star-forming galaxies at $z>2.7$ to be effectively detectable in the
ultraviolet bandpass. 

\sectionbegin{3. THE COLORS OF HIGH REDSHIFT GALAXIES}

One of the major objectives of the HDF observations is to determine the
morphology, spectral energy distribution, and number density of 
galaxies actively forming stars at high-$z$, the nature of which is crucial to
our understanding of galaxy evolution. In this section we shall show how we can
use intergalactic absorption to identify high redshift galaxies in broadband
multicolor surveys.  Colors can be more readily interpreted using the
AB-magnitude system (Oke 1974)

$$m_{AB}=-2.5\log \int f(\nu)e^{-\tau_{eff}}T(\nu)d\nu -48.6, 
\eqno(\new)$$ 

\ni where $f(\nu)e^{-\tau_{eff}}$ is the mean incident power measured in
$\funits$, and $T(\nu)$ is the normalized transmittance of the relevant filter.
We put our magnitudes directly into the AB system, so that
$U_{300}=B_{450}=V_{606}= I_{814}$ corresponds to a spectrum for which
$f(\nu)e^{-\tau_{eff}}$ is constant. The observed colors of distant objects are
determined by the intrinsic spectral energy distributions (SEDs) of their
stellar populations, absorption and scattering of photons in their interstellar
media, and stochastic absorption due to intervening \HI clouds. Because we do
not know when galaxies started forming stars, and over what timescale they
formed them, it is impossible to predict their energy distribution in detail.
Furthermore, it is not necessarily the case that the faintest galaxies are the 
most distant. Thus our analysis must not rely on fluxes or apparent
magnitudes to give redshift information.
It is, however, possible to outline in a general way the range of plausible
colors for star-forming galaxies at high redshift, and to identify combinations
of age and dust reddening in a stellar population at lower redshift that may
conspire to mimic a high-$z$ galaxy. 

\subsectionbegin{3.1. The Spectral Energy Distribution of Star-Forming Galaxies}

For illustrative purposes, we analyze here two simple examples in detail: 

(1) We compute the ultraviolet spectrum of a star-forming galaxy using the
isochrone synthesis spectral evolutionary code of Bruzual \& Charlot (1993).
The model has solar metallicity, a constant star formation rate at age 0.3 Gyr,
a Salpeter (1955) initial mass function (IMF), $\phi(m)\propto m^{-2.35}$, with
lower and upper cutoffs of 0.1 and 125$\msun$, and completely ignores the
effects of \HI in the local interstellar medium on the transfer of ionizing
photons. Two intrinsic discontinuities, about 1.4 mag across the rest-frame
Lyman limit and 0.5 mag across the Balmer decrement, are characteristic of the
most prominent stars; 

(2) We take as fiducial the spectra of the four brightest starburst galaxies
observed by the {\it Hopkins Ultraviolet Telescope} (HUT) on Astro-2: NGC3310,
4214, 5236 and 5253 (Ferguson \etal 1996a). The {\it HUT} spectra cover the
wavelength range $912<\lambda<1820\,$\AA\ (Kruk \etal 1995). The cutoff at
short wavelengths is due to galactic \HI. For wavelengths longward of
$1820\,$\AA, we have patched on the {\it IUE} spectra from the Kinney \etal
(1993) atlas, renormalized to the {\it HUT} flux in the range $1350-1700\,$\AA.
The renormalization is required because of the different sizes of the {\it HUT}
and {\it IUE} apertures.  Note that the starburst galaxy intrinsic colors are
significantly redder than the ones derived from the population synthesis model
because of the presence of a large Lyman discontinuity and of dust-reddening.
{\it HUT} spectra of somewhat more distant starbursts galaxies suggest that the
intrinsic Lyman break is typically at least  a factor of 5 (Leitherer
\etal 1995). Analysis of the NGC5236 {\it HUT} SED suggests that a young
($\sim 5\times 10^6$ year old) population is responsible for the emission, with
an intervening extinction $E(B-V)\approx 0.2$ necessary to fit the detailed SED
(Ferguson \etal 1996a). Such a modest extinction appears fairly typical
of starburst galaxies (even those of high metallicity). In a comprehensive
study of 39 starburst galaxies, Calzetti, Kinney, \& Storchi-Bergmann (1994) 
found values of $E(B-V)$ ranging from 0 to 0.8.

Figure 2{\it a} displays the predicted $U_{300}-B_{450}$, $B_{450}-V_{606}$, and
$V_{606}-I_{814}$ colors versus emission redshift of our synthetic 
IGM-attenuated galaxy, together with the unattenuated values. The intrinsic 
spectral shape has been kept constant with cosmic time. As expected, the observed colors 
redden quite sharply with redshift due to intergalactic absorption. 
For example, from $z_{\rm em}=2.5$ to $z_{\rm em}=3$, the observed
$U_{300}-B_{450}$ color of the Bruzual spectrum increases from 1.8 to 3.6 mag.
In the absence of any absorption break, the last value would correspond to a
spectral energy distribution $f(\nu)\propto \nu^{-\alpha}$ with
$\alpha=8$.\note{Above $z_{\rm em}\approx3.5$, $U_{300}-B_{450}$ becomes bluer,
again an effect of the red-leak of the F300W filter. However, by then,
intergalactic attenuation is so strong that high-$z$ galaxies are effectively
unobservable in the ultraviolet.}~ The observed flux decrements between 4500
and $6060\,$\AA\ and between 6060 and $8140\,$\AA\ similarly probe the universe
at higher and higher redshifts, as $B_{450}-V_{606}$ increases from 0.8 at
$z_{\rm em}=3.5$ to 1.7 mag at $z_{\rm em}=4$, while $V_{606}-I_{814}$
increases from 1.0 at $z_{\rm em}=4.5$ to 1.9 mag at $z_{\rm em}=5$. 
A similar plot is shown in Figure 2{\it b} for the observed {\it HUT-IUE}
starburst galaxy spectra. The main point to note here is how intrinsically blue
and red objects follow similar paths in the color-redshift plane, and that,
even in the presence of dust-reddening, {\it the large magnitude jumps between
adjacent bandpasses are largely caused by the known sources of intergalactic
\HI opacity -- the \Lya clouds and Lyman-limit systems -- rather than by the
assumed galaxy SED.} 

\subsectionbegin{3.2. Color Selection Criteria}

The problem of identifying high redshift galaxies is complicated 
by the fact that the present-day colors and spectra of galaxies allow
for a very wide variety of star-forming histories. This evolution 
is coupled with chemical evolution of the stars and the ISM 
and changes in the amount of dust and its distribution relative to the stars.
However, galaxy spectra are not entirely arbitrary. Flux is usually a slowly
varying function of wavelength, with a few spectral discontinuities (e.g., the
4000 and $912\,$\AA\ breaks) at specific locations. Thus it should be possible
to construct fairly robust selection criteria that will exploit the combined
effect of the intrinsic Lyman edge in galaxies and the opacity of intergalactic
neutral hydrogen to separate high redshift from low redshift objects. 

In this section, we shall define criteria appropriate to the HDF bandpasses. To
derive a robust color selection technique, we have computed HDF colors for 1612
synthetic spectra of galaxies, representing different ages, star formation
histories, metallicities, and dust opacities. The stellar-evolutionary input to
the models has been described in detail by Babul \& Ferguson (1996). Briefly,
synthetic spectra are computed using isochrone synthesis from the isochrones of
Bertelli \etal (1994), and the model atmospheres of Kurucz (1992) and Clegg \&
Middlemass (1987). The synthetic spectra agree well with those of Bressan,
Chiosi, \& Fagotto (1994), which were constructed using the same isochrones,
and slightly less well with those of Bruzual \& Charlot (1993), primarily due
to their use of different isochrones. The uncertainties in population synthesis
have been outlined in some detail by Charlot, Worthey, \& Bressan (1996). These
uncertainties amount to differences in color of several tenths for the same
stellar population computed by different codes. While they can cause very large
discrepancies in the derived ages, such small variations in color are almost
completely negligible for our purposes. 

Attenuation by dust has been included using the parametrized extinction laws of
Pei (1992) for the Galaxy, the LMC and the SMC. Note that the use of the 
relatively ``gray'' extinction curve for starburst galaxies found by Calzetti 
\etal (1994) would imply a higher UV continuum than predicted on the basis of
the optical continuum emission by the application of the standard extinction
laws, hence strengthen the efficiency of our selection criteria. We ignore the
effects of \HI in the local interstellar medium of the galaxies, which will add
to the intergalactic attenuation to make high-$z$ objects even redder. There
are no emission lines included in the simulations.  Given the very wide
bandpasses of the HDF filters, we do not expect emission lines to significantly
alter galaxy broadband colors. 

The colors for the models have been computed at 73 different redshifts spanning
the interval $0.001<z<7$, taking into account the effect of intergalactic
attenuation. In this exercise, galaxies are required to be younger than the age
of the universe (for $q_0=0.5, H_0=50 \rm\, km\,s^{-1}\,Mpc^{-1}$), but
otherwise age has been decoupled from redshift. The model grid is intended
to span the range of plausible colors of real galaxies, but not to sample this
range in a way that is tied to cosmological models. For example, in the real
universe we expect the old, low redshift portion of a color-color diagram 
to be more densily populated than in our simulations. The
adopted grid of ages, star formation timescales, extinctions $A_B$, and
redshifts is summarized in Table 1. It is important to emphasize that our
goal is to construct a {\it robust} set of color selection criteria that
are {\it largely independent} of models of galaxy formation or the
assumed values of the cosmological parameters. In this respect our approach
is quite distinct from efforts to assign photometric redshifts to each
galaxy, as presented by Lanzetta, Yahil, \& Fern{\' a}ndez-Soto (1996)
and Gwyn \& Hartwick (1996), for example.

\subsubsectionbegin{3.2.1. F300W Dropouts}

We are now in position to examine the colors of arbitrary galaxies as a
function of redshift. Figure 3 shows $\ub$ vs. $\bi$ for all of the synthetic
spectra. The large points show objects with $2<z<3.5$.
Galaxies in this redshift range predominantly occupy the top left portion of
the plot because of the attenuation by the IGM and intrinsic
extinction. Galaxies at lower redshift can have similar $\ub$ colors, but they
are typically either old or dusty, and are therefore red in $\bi$ as well. To
identify likely star-forming galaxies in our desired redshift range, we require
that they have (1) $\ub>1.3$, (2) $\ub>\bi+1.2$, and (3) $\bi<1.5$. These
criteria isolate objects that have relatively blue colors in the optical, but a
sharp drop into the UV. 

Figure 4 gives an estimate of the efficiency of this technique. The
solid histogram shows the fraction of the total sample of models that meet the
selection criteria as a function of redshift. Even with the wide range of
parameters in our galaxy models, {\it no} galaxies with $z<1.5$  or $z>4$ are
selected. However, many of the galaxies in the redshift range $2<z<4$ are
missed as well. The galaxies that are missed tend to be redder than the
selection line in $\bi$, either because they are relatively old, or because
they are highly reddened by dust. The dotted histogram in Figure 4 shows the
selection efficiency for galaxies with ages restricted to be less than $10^8$
yr and extinctions $A_B<2$. The selection criteria appear to be extremely
efficient, recovering roughly 90\% of the galaxies in the redshift range $2<
z<3.5$. Of course the true efficiency depends on how close our models are to
real protogalaxies, and on how many old or very dusty galaxies there are in the
real universe in that redshift range. 

\subsubsectionbegin{3.2.2. F450W Dropouts}

We can play an identical game with the F450W dropouts. Figure 5 shows the
colors of the model galaxies in the $\bv$ vs. $\vi$ plane, with galaxies in the
redshift range $3.5<z<4.5$ highlighted, and with our proposed selection
criteria drawn. In this case, there are objects outside the redshift range of
interest that can have rather blue $\vi$ colors and red $\bv$ colors, and hence
potentially contaminate the sample. These peculiar colors obtain only for those
galaxies that lie in the redshift range $ 2<z<3.5$ and have a fair amount
of internal extinction. The blue $\vi$ color is due to the $2200\,$\AA\ 
absorption feature shifting into the F814W bandpass. Our selection criteria are
tuned to avoid including many of these galaxies. Specifically, we require our
candidates to have (1) $\bv>1.5$, (2) $\bv>1.7(\vi)+0.7$, (3)
$\bv<3.5(\vi)+1.5$, and (4) $\vi<1.5$. 

Figure 6 shows the fraction of galaxies recovered as a function of redshift
for such a selection. Once again, the criteria appear to be very efficient at
selecting relatively unreddened star-forming galaxies, this time in the
redshift range $3.5 < z < 4.5$. Neverthless, there is some danger of
low redshift contamination for the F450W dropouts. Again, most of the
contaminants are either relatively old or reddened galaxies at slightly lower
redshift. The dashed histogram in Figure 6 shows the distribution of redshifts
for the sample when galaxies are restricted to have either ages greater than
$10^9$ yr, or $A_B > 2$. Again, the extent of the contamination in the real
universe depends on the unknown distribution of dust contents and ages at these
redshifts. 

\sectionbegin{4. HIGH REDSHIFT GALAXIES IN THE HUBBLE DEEP FIELD}

Using the color selection criteria described above, we shall build in this
section a sample of galaxies at high-$z$ selected from the Version 2 catalog of
the {\it Hubble Deep Field}, which is based on the second release of the
reduced HDF images. Details of the data reduction, image combination, source
detection, and photometry are given by Williams \etal (1996). 

Figures 7{\it a} and 8{\it a} show color-color plots of the catalog data with 
our selection region drawn in as dashed lines.  The catalog is based on 
the three WF chips
only, and covers an area of 4.65 arcmin$^2$. The source detection was
carried out on the exposure-weighted sum of the F606W and F814W images, and the
threshold was set to $4\sigma$. In all cases, isophotal magnitudes and colors
were used for photometry, with identical apertures (defined from a sum of the
F606W and F814W images) applied to the images in all bandpasses. It is
important to note that the FOCAS isophotal magnitudes are systematic
{\it underestimates} of the true brightness of the galaxies. The magnitude
error is difficult to quantify precisely as it depends on the surface brightness
profiles of the galaxies in the image. Experiments on simulated data suggest
that total magnitudes are typically 0.5 mag brighter than FOCAS isophotal
magnitudes (Ferguson \etal 1996b). We have ignored this offset in our analysis,
but note that it could increase our inferred luminosity densities and metal
production rates by roughly a factor of 1.6. The catalog
was truncated at magnitude limits sufficiently bright that Lyman breaks of the
expected amplitude could be measured reliably.  For the $UBI$ selection, we
considered only objects with $B_{450} < 26.79$ and $V_{606} < 28.0$;  for the
$BVI$ selection, the limiting threshold was set to $V_{606} < 27.67.$ Galaxies
undetected in F300W (F450W) -- as defined by having signal-to-noise ratio $<1$
inside the isophotal aperture -- were assigned a $1\sigma$ lower limit to their
$U_{300}-B_{450}$ ($B_{450}-V_{606}$) colors.  In the figures, only objects
with $I_{814} > 21$ are shown.  For the $UBI$ selection of F300W dropout
objects, two bright stars (with $I_{814} < 21$) enter into the color selection
window. Steidel \etal (1996a) have shown that subdwarf stars may have similar
colors to those of $z \sim 3$ galaxies, and indeed Steidel \etal (1996b)
confirm that one of the two bright stars in the HDF within this color selection
window is indeed a foreground subdwarf star.  None of the fainter F300W
dropouts appears to be stellar. 

The complex, sometimes multi-component morphologies of faint galaxies in
deep {\it HST} images provide a complication for anyone cataloging faint
objects in images such as the HDF.  Automated object detection routines
may break up objects into several components if their light distribution
has multiple peaks, and it is not always clear whether a multi-component
object should be considered as a single ``parent'' entity or split into 
separate ``daughters.'' We chose to apply our selection criteria to 
the complete catalog of both parents and daughters, and then inspected
all potential high redshift candidates visually.  In a few cases, we retained
``parents'' as single objects (e.g., for the brightest F300W dropout,
the elongated object known as C4--06 in the notation of
Steidel \etal 1996b), while for others we chose the ``daughters'' when
the parent seemed to comprise distinctly separate objects.  
Although we shall consider here only the total, summed light from all candidate 
Lyman break objects, the details of this splitting can be of some importance
as the daughters may fall in or out of our sample.

Ultimately, of course, the ability of our color criteria to select 
largely uncontaminated samples of star-forming galaxies at high-$z$
must be confirmed by deep spectroscopy. 
In Figures 7{\it b} and 8{\it b} we plot the location in the color-color 
diagrams of the more than 60 galaxies in the HDF which have known Keck/LRIS
spectroscopic redshifts (Steidel \etal 1996b; Cohen \etal 1996; Moustakas
\etal 1996). While none of the low-$z$ objects is found to contaminate
our high-$z$ selection regions, our $UBI$ color criteria are able to
select 4 of the 5 objects spectroscopically confirmed by Steidel
\etal (1996b) to be galaxies in the redshift range $2.591<z<3.226$. (Here,
photometry from the Version~2 catalogs has been used, whereas 
Steidel \etal used Version~1 data.)  Two out of the six known $z>2$ objects 
in the HDF are {\it not} selected by our criteria.
One (a $z=2.268$ galaxy from the Cohen \etal sample) only just misses being
included -- its $U-B$ color is slightly bluer than allowed by our
color criteria. The second missed source (C4--09 from Steidel
\etal) is a more puzzling case -- this object is a small diamond-shaped
configuration of four distinct sub-components, each of which ``disappears'' in
the F300W bandpass. Steidel \etal considered it as a single object and measured
a redshift of 3.23.  We fail to identify it as a candidate because it appears
as a formal detection (2.7$\sigma$) at F300W in the Version 2 catalog, whereas
Steidel \etal assigned only an upper limit to its $U_{300}$ flux. This points
to some uncertainty in our current understanding of the noise properties of the
F300W images, but we do not expect this to significantly impact the results
presented here.  In the Version 2 catalog, the brightest of the four
sub-components of this small ``quad'' is individually classified as a dropout
above our magnitude limit and with the appropriate colors, and is included in
the analysis presented here. 

\subsectionbegin{4.1. F300W Dropouts: Limits to the Luminosity Density 
at $2<z<3.5$}

We have identified 69 F300W dropouts in the HDF images which satisfy the
criteria established for star-forming galaxies at $2<z<3.5$. Assuming this
redshift interval has been uniformly probed, we derive a comoving galaxy number
density of $4.2\times 10^{-3}\,$Mpc$^{-3}$ ($0.92\times 10^{-3}\,$Mpc$^{-3}$
for $q_0=0.05$) at $\langle z\rangle=2.75$, about 10 times larger than the
comoving space density of bright $L\ge L_*$ present-day galaxies (Loveday \etal
1992). The observed $V_{606}$ magnitudes of our sample yield a specific
(comoving) emissivity at $1620\,$\AA\ equal to \eqnam{\eud} 
$$
\rho_{1620}\approx 1.6\times 10^{26}\sunits \eqno(\new)
$$
($9.0\times 10^{25}\sunits$ for $q_0=0.05$). 
The faintest F300W dropout has a magnitude of $V_{606}=26.7$, corresponding, in
the case of a flat-spectrum source at $\langle z\rangle=2.75$, to a B-band
luminosity of about 0.1$L_*$, or a total star formation rate
(Salpeter IMF) of $\sim 1\msun$ yr$^{-1}$. The brightest has $V_{606}=23.6$,
and is forming stars at a rate of $\sim 20\msun$ yr$^{-1}$.

It is interesting to compare the comoving space density of the HDF F300W
dropouts brighter than $V_{606}=25$, $7.3\times 10^{-4}\,$Mpc$^{-3}$ at
$\langle z\rangle=2.75$, with that derived from Steidel \etal (1996a)
ground-based statistic, $3.6\times 10^{-4}\,$Mpc$^{-3}$ to ${\cal R}<25$ at
redshift $\langle z\rangle=3.25$. Within the errors, the two estimates 
appear in good agreement with one another, especially after accounting for 
the fact that one probes $\sim 0.3$ mag fainter in the galaxy luminosity
function at the HDF average redshift.

\subsectionbegin{4.2. F450W Dropouts: Limits to the Luminosity Density 
at $3.5<z<4.5$}

In a similar manner, we have identified 14 F450W dropouts which satisfy the 
criteria established for star-forming galaxies at $3.5<z<4.5$. Their
comoving number density at $\langle z\rangle=4.0$ is $1.5\times 
10^{-3}\,$Mpc$^{-3}$ ($2.4\times 10^{-4}\,$Mpc$^{-3}$ for $q_0=0.05$). 
From the observed $I_{814}$ magnitudes, we derive a
specific (comoving) emissivity at $1630\,$\AA\ of \eqnam{\ebd} 
$$
\rho_{1630}\approx 5.0\times 10^{25}\sunits \eqno(\new)
$$
($2.8\times 10^{25}\sunits$ for $q_0=0.05$). 
The faintest F450W dropout has a magnitude of $I_{814}=27.5$, corresponding,
in the case of a flat-spectrum source at $\langle z\rangle=4$, again 
to a B-band luminosity of about 0.1$L_*$, while the brightest has $I_{814}=25$.
Our samples of $U_{300}$ and $B_{450}$ dropouts therefore reach comparable
depths. {\it We stress that, contrary to the case at $2<z<3.5$, we have yet no
spectroscopic confirmation of the efficiency of our color-selection criteria 
at $3.5<z<4.5$.}

The source catalogs of the $U_{300}$ and $B_{450}$ dropouts are presented
in Tables 2 and 3. For each galaxy we report the FOCAS catalog entry number
from Williams \etal (1996) (ID), the right ascension and declination ($\alpha,
\delta$), and the AB isophotal magnitudes in the four HDF passbands.\note{The
use of a more conservative criterion which assigns a $2\sigma$ lower 
limit to the $U_{300}-B_{450}$ colors of galaxies undetected in the F300W
bandpass would decrease the number of ultraviolet dropouts to 47, thereby
decreasing the integrated emissivity given in eq. (\eud) by a factor of
1.3. The same $2\sigma$ color criterion applied to galaxies undetected in 
F450W decreases to only 9 the number of blue dropouts, resulting in an
emissivity which is a factor 1.4 lower than that given in eq. (\ebd).}~
The UV dropouts 2-449.0, 3-550.0, 4-555.1, and 4-676.0 have been 
spectroscopically identified by Steidel \etal (1996b) as galaxies at 
$z=$2.845, 2.775, 2.803, and 2.591, respectively.

\sectionbegin{5. COSMIC STAR FORMATION HISTORY}

The results of the Canada-France Redshift Survey (CFRS, Lilly \etal 1995) over
the redshift range $0<z<1$, the spectroscopic confirmation by
Steidel \etal (1996a,b) of the existence of a substantial population of
luminous, star-forming galaxies at $2.5<z<3.5$, together with our statistical
constraints on the redshift distribution of faint galaxies in the {\it Hubble
Deep Field}, as derived above for $2<z<4.5$, have considerable implications
for our understanding of the global history of star and structure formation in
the universe. Here, we review some of the consequences of these observations,
focusing on what can be learned about galaxy evolution at early (and late)
epochs from integrated quantities over the entire population, rather than from
a detailed study of individual objects. 

\subsectionbegin{5.1. The Metal Production Density of the Universe}

The UV continuum emission from a galaxy with significant ongoing star formation
is totally dominated by short-lived massive stars, and is therefore nearly
independent of the galaxy history. Moreover, the (rest-frame) radiation flux
below $3000\,$\AA\ is a very good measurement of the instantaneous ejection
rate of heavy elements ($Z\ge 6$), since both are directly related to the
number of massive stars (Cowie 1988; Songaila, Cowie, \& Lilly 1990; Madau \&
Shull 1996): the same stars with $m>10\msun$ that manufacture and return most of
the metals to the ISM also dominate the UV light.  The supernova
event from a star of $25\msun$ injects about $4.5\msun$ of metals (Woosley \&
Weaver 1995). At the end of the C-burning phase
$\sim 17\msun$ have been converted into helium and carbon (Maeder
1992), with a mass fraction released as radiation of 0.007. For each
$1\msun$ of metals ejected, to first approximation we then expect
$(17\times 0.007/4.5)=0.025\msun$ c$^2$ of energy to be radiated away.

What we are interested in here is the universal rate of ejection of newly
synthesized material per unit comoving volume. In the approximation of
instantaneous recycling, the metal ejection rate (MER) per unit volume can be
written as (cf Tinsley 1980) \eqnam{\mer} 
$$ 
{\dot \rho_Z}=\psi\int mp_{\rm zm}\phi(m)dm, \eqno(\new)
$$
where $\psi$ is the star formation rate (SFR) density, $\phi(m)$ is the IMF
(normalized through the relation $\int m\phi(m)dm=1$), and $p_{\rm zm}$ is the
stellar yield, i.e., the mass fraction of a star of mass $m$ that is converted
to metals and ejected. The dot denotes differentiation with respect to cosmic
time. At short wavelengths, the luminosity density radiated per unit frequency
during the main sequence phase is related to $\psi$ by 
$$
\rho_\nu=0.007c^2\psi\int m f_{\rm He}(m)f_\nu(m)\phi(m)dm, \eqno(\new)
$$
where $f_{\rm He}(m)$ is the mass fraction of hydrogen burned into helium, and 
$f_\nu(m)$ is the normalized spectrum of stars of mass $m$ on the main sequence.
The detailed conversion from UV luminosity to MER, including all evolutionary
stages, can be obtained from spectral synthesis and metal-yield calculations.
For simplicity, we shall ignore here the contributions of planetary nebulae
(Renzini \& Voli 1981), Type Ia supernovae (Thielemann, Nomoto, \& Yokoi 1986),
and stellar winds (Maeder 1992) to the chemical yields, and focus on the
nucleosynthetic enrichment by supernova explosions from massive stars. We use
the evolutionary models of Bruzual \& Charlot (1993) and the Type II stellar
yields tabulated by Woosley \& Weaver (1995) (see also Sutherland \& Shull
1996). For a Salpeter IMF including stars in the $0.1<M<125\msun$ mass range, a
constant SFR, a galaxy age in the interval 0.1--1 Gyr, and solar metallicity,
we derive \eqnam{\bcone} 
$$
\rho_{1500}=4.4\pm 0.2 \times 10^{29} {\dot \rho_Z}~ \sunits \eqno(\new)
$$
at $1500\,$\AA, and \eqnam{\bctwo} 
$$
\rho_{2800}=2.8\pm 0.3\times 10^{29} {\dot \rho_Z}~ \sunits \eqno(\new)
$$
at $2800\,$\AA, where $\dot \rho_Z$ is the MER in $\msunits$. The continuum
spectrum is fairly flat and is dominated by early-type stars. We shall use 
the relations given above in the subsequent discussion, while noting that the
conversion efficiency is fairly insensitive to the assumed IMF (as long as the
stellar population extends as a power-law to massive stars, 50--100
$M_{\odot}$), since the increased metal yield from high mass stars is
compensated for by a similar increase in the production of UV photons. The
advantage then of deriving from the observed UV luminosity density a rate of
metal ejection, rather than a SFR, is that the latter is instead a sensitive
function of the IMF slope. To be quantitative, while for a Salpeter IMF the
mean yield of returned metals is 
$$ 
\int mp_{\rm zm}\phi(m)dm=2.4\%, \eqno(\new)
$$
for a Scalo (1986) IMF -- less rich in massive stars -- in the same mass range
this factor is about 3.3 times lower. Hence large errors in the relative rates
of star formation at different epochs may occur if the IMF varies with cosmic
time. Because of these considerations, unless otherwise stated we shall only
quote metal ejection rates in the following. Note, however, that the real
uncertainties in the UV-to-metal conversion factors due to age, IMF,
metallicity, and population synthesis model well exceed the errors quoted
above. 

To relate the amount of heavy elements observed today in various stellar
populations and in the gas phase to the UV luminosity of distant, star-forming
galaxies, we may use the conservation of metals, which implies that 
the sum of the heavy elements stored in stars, $Z_*\rho_*$, and in the
gas, $Z_g\rho_g$, is equal to the mass of metals ever ejected, i.e., to the 
integral of equation (\mer) over cosmic time. An estimate of the baryonic mass
in galaxies can be obtained by multiplying the observed luminosity density of
the local universe by a mass-to-light ratio. Since the gaseous content of
nearby galaxies is observed to be negligible compared with the stellar content
(e.g., Rao \& Briggs 1993), the cosmological mass density of heavy elements at
the present epoch is given by 
$$
\rho_Z(0)\approx Z_*\rho_*(0)=Z_*\rho_B(0)({M\over L_B}), \eqno(\new)
$$
where $\rho_B(0)$ is the local blue light density of field galaxies, and $M/L_B$
is the mass-to-blue light ratio of visible matter. There are several recent
determinations of the B-band luminosity density from large redshift surveys.
The values obtained for $\log \rho_B(0)$ in units of L$_\odot\,$Mpc$^{-3}$ are
7.98 (Efstathiou, Ellis, \& Peterson 1988), 7.83 (Loveday \etal 1992), 8.0
(Marzke, Huchra, \& Geller 1994), 7.93 (da Costa \etal 1994), and 8.0 (Ellis
\etal 1996). A simple average of all these determinations gives the adopted
value: 
$$
\rho_B(0)=9.0\pm 1.4\times 10^7\,{\rm L_\odot\, Mpc^{-3}}. \eqno(\new)
$$
The mass-to-light ratios for the visible parts of galaxies can be estimated
from stellar population modeling. The results (in solar units) are in the
range $2<M/L_B<6$, the smaller value referring to a typical Population I system
like the solar neighborhood (e.g., Kuijken \& Gilmore 1989), the larger value
to the old stellar populations of elliptical galaxies (e.g., van der Marel
1991). Since about 30\% of the total blue luminosity density is in ellipticals
(Efstathiou \etal 1988), we may adopt the luminosity-weighted mean value 
of $\langle M/L_B\rangle=3$ to derive \eqnam{\m} 
$$
\rho_Z(0)\approx 5.4\pm 0.8\times 10^6 ({Z_*\over Z_\odot})\munits
\eqno(\new) 
$$ 
(cf Cowie 1988). While representative stellar metallicities in nearby luminous
spirals and ellipticals are approximately solar, $Z_*\sim Z_\odot=0.02$,
irregular, blue compact, and dwarf spheroidal galaxies are known to be metal
poor, $Z_*\gta 0.1Z_\odot$ (e.g., Pagel \& Edmunds 1981). 

The real uncertainty on the metal mass density is difficult to
estimate, but is unlikely to exceed a factor of 1.5. Although a baryonic 
mass several times larger than the luminous mass may be present in the 
Galactic halo, metal-rich halo material would be mixed into and
over-enrich the disk. Hence, if a substantial amount of metals are missing from
our census, they are most likely hidden in low-surface brightness galaxies
(Ferguson \& McGaugh 1995), or intergalactic and intracluster gas. 

It is useful at this stage to define a fiducial metal ejection density,
\eqnam{\fid} 
$$
{\dot \Sigma_Z}=\rho_Z(0)t_H(0)^{-1}\approx 4.2\times 10^{-4}\msunits, 
\eqno(\new)
$$
given by the present-day mass density in heavy elements divided by the current
age of the universe. If ${\dot \rho_Z} $ observed at a given redshift
$z$ is much less than ${\dot \Sigma_Z}$, and a large fraction of the luminous
baryons observed today were already locked into galaxies at this epoch, then
either galaxies have already exhausted their reservoirs of cold gas or there
must be a mechanism which prevent the gas within virialized dark matter halos
to radiatively cool and turn into stars. By contrast, a metal ejection rate
much greater than the fiducial value implies that the conversion of gas into
stars can be extremely efficient at times. From a comparison between the
observed ${\dot \rho_Z}$ and ${\dot \Sigma_Z}$ we may then be able to
distinguish between a more or less uniform (with cosmic time) SFR and 
a discontinuos one, consisting of a series of bursts of suitable duty cycle. 

According to Gallego \etal (1995), the \Ha luminosity density of the local
universe is $\rho_{\rm H\alpha}(0)=1.3\pm 0.6\times 10^{39}\ldunits$. This
value implies\note{In case-B recombination theory, the \Ha luminosity density
can be related to the emission rate of ionizing photons per unit cosmic volume,
$\dot n_{\rm ion}$, according to $\rho_{\rm H\alpha}\approx 0.45h\nu_{\rm
H\alpha}\dot n_{\rm ion}$ (Osterbrock 1989). Population synthesis galaxy
spectra (Bruzual \& Charlot 1993) with constant SFR yield the following
approximate relation between $\rho_{1500}$ and $\dot n_{\rm ion}$: $\dot n_{\rm
ion}\approx 0.14\rho_{1500}/h$, nearly independent of age in the interval
0.1--1 Gyr.}~ a MER at the present epoch of \eqnam{\mpr} 
$$
{\dot \rho}_Z(0)\approx 1.1\pm 0.5\times 10^{-4}\msunits. \eqno(\new)
$$
Equations (\fid) and (\mpr) suggest that the present rate of
production of heavy elements is too low to yield the observed element
abundances in the Hubble time, and that either star-forming galaxies were much
more numerous in the past, or individual galaxies must, on average, have passed
through a significantly brighter phase. 

From the CFRS faint galaxy survey (Lilly \etal 1995), we now know that the MER 
per unit mass was in fact considerably larger in the past. The comoving metal
ejection density at $0<z<1$ corresponding to the observed specific luminosity
density at $2800\,$\AA\ (the ``LF-estimated'' value of Lilly \etal 1996) is
\eqnam{\mprr} 
$$
{\dot \rho}_Z(z)\approx 1.5\pm0.6\times 10^{-3}\left({1+z\over
2}\right)^{3.9} \msunits, \eqno(\new) 
$$
about 15 times higher at $z\sim 1$ than the current value. This strong
evolution is associated with galaxies bluer than typical Sbc's. The luminosity
function of redder galaxies shows little change back to $z\sim 1$.

The sample of Lyman-break star-forming galaxies identified by Steidel \etal
(1996a) has a comoving density of $\approx 3.6\times 10^{-4}\,$Mpc$^{-3}$ at
$\langle z\rangle=3.25$, and an average continuum specific luminosity at
$1500\,$\AA\ of $10^{41}\,$ergs s$^{-1}$ \AA$^{-1}$. This gives a comoving
volume emissivity of $2.7\times 10^{25} \sunits$, hence a MER density 
$$
{\dot \rho}_Z(3.25)\approx 6.2\times 10^{-5}\msunits, \eqno(\new)
$$
comparable with the local value but more than 20 times lower than the $z=1$ one.
\note{Note that the star formation density at $z=0$ quoted by Gallego \etal
(1995) is derived assuming a Scalo IMF, while Steidel \etal (1996a) adopt a 
Salpeter IMF in their paper.}

Finally, our analysis of the HDF images yields, from equations (\eud)
and (\ebd): \eqnam{\hdfone}
$$
{\dot \rho}_Z(2.75)\approx 3.6\times 10^{-4}\msunits, \eqno(\new)
$$
and \eqnam{\hdftwo}
$$
{\dot \rho}_Z(4)\approx 1.1\times 10^{-4}\msunits. \eqno(\new)
$$
Note that the MER density at $\langle z\rangle=2.75$ is more than a factor of 5
larger than the Steidel \etal (1996a) ground-based value at $\langle z
\rangle=3.25$, largely because the HDF deep images probe about 1.7~mag
fainter into the rest-frame luminosity function at these redshifts. Still, 
the derived luminosity density is significantly smaller than the rate at
$z\approx 1$. It is conceivable that our list of candidates in the HDF may be
sampling a significant fraction of the galaxy luminosity function at these
redshifts, and that the contribution to the luminosity density from galaxies
below our magnitude threshold is small. Strictly speaking, however, the metal
ejection rates derived above from the HDF data (and the emissivities given in
eqs. [\eud] and [\ebd]) should be interpreted as lower limits to the real
values, as they only include the fraction arising in the most actively
star-forming, young and nearly dust-free objects. The same considerations also
apply to the sample identified by Steidel \etal (1996a). We point out that
large numbers of faint galaxies at high redshifts are quite plausible in cold
dark matter-like cosmologies, because of the high space density of low mass
dark matter halos predicted in any hierarchical theory of galaxy formation
(e.g., White \& Frenk 1991). 

The derived star and element formation history of the universe is depicted in
Figure 9. Although the star formation densities beyond $z=2$ are only lower
limits, together with the Gallego \etal and Lilly \etal values they seem
consistent with the existence of a peak in the cosmic metal production rate in 
the redshift range $1\lta z\lta 2$, as predicted by models of the chemical
evolution of the damped \Lya absorption systems (Pei \& Fall 1995). 

\subsectionbegin{5.2. Clues to Galaxy Formation and Evolution}

We may at this stage try to establish a cosmic timetable for the production of
heavy elements in relatively bright galaxies, keeping in mind the inherent
uncertainties associated with the estimates given above. In an Einstein-de
Sitter universe, the total mass density processed into metals over the redshift
range $0<z<z_c\approx 1$, is, integrating equation (\mprr) over cosmic time,
\eqnam{\mm} 
$$
\Delta\rho_Z(0)=(8.2\pm 2.6\times 10^5\munits) [(1+z_c)^{2.4}-1]=
3.5\pm 1.1\times 10^6\munits. \eqno(\new)
$$
(Contrary to the measured number densities of objects and rates of star
formation, the metal mass density does not depend on the assumed cosmological
model.) If we define two characteristic epochs of star and element formation in
galaxies, $z_*$ and $z_Z$, as the redshifts by which half of the current
stellar and metal content of galaxies was formed, then a straightforward
comparison between equations (\m) and (\mm), together with the fact that most
of the stars in the inner luminous parts of galaxies are metal rich, imply
$z_*\lta z_Z\approx 1$, or in other words that a significant fraction of the
current metal content of galaxies was formed relatively late, on a timescale of
about 8 Gyr. This is comparable with the decay time of star formation, from 5
to 9 Gyr, required to reproduce the present-day colors of late-type spirals
(e.g., Bruzual \& Charlot 1993), and suggests the possibility that we may be
observing in the redshift range $z=0-1$ the conversion into stars of gaseous
galactic disks. Pure \HI disks may be assembled at some higher redshift, and
disk gas continuosly replenished as a result of ongoing infall from the
surrounding hot halo. The case for a ``disk epoch'' at $z\sim 1$ is
strengthened by the observations that galaxies of different luminosities and
morphological type evolve very differently. A common conclusion of several
recent deep redshift surveys (e.g., Glazebrook \etal 1995; Lilly \etal 1995;
Ellis \etal 1996) is that late-type, gas-rich systems show substantial
evolution in number and/or luminosity from the present to $z\sim 1$, and 
that a large fraction of the observed evolution is associated with an increase
in the mean surface brightness of luminous disks (Schade \etal 1995). By
contrast, the ellipticals and early-type spirals have been remarkably quiescent
over the same redshift interval. 

From stellar population studies we also know 
that about half of the present-day stars -- hence metals -- are contained into
spheroidal systems, i.e., elliptical galaxies and spiral galaxy bulges, and
that these formed early and rapidly. The major arguments come from studies of
Galactic bulge color-magnitude diagrams (Ortolani \etal 1995) which indicate
ages as old as globular clusters, and from studies of the chemical abundances
in cluster ellipticals and cluster X-ray gas, whose high $\alpha$-element
ratios suggest that both the stars and the intracluster gas must have been
enriched on timescales of order 1 Gyr or less (Renzini \etal 1993). If these
arguments are correct (and they are by no means undisputed), then bulges and
ellipticals must have experienced a bright starburst phase at high $z$. Where
are these protospheroids?

There are a few circumstantial pieces of evidence supporting the interpretation
that the Lyman-break galaxies identified at $z>2.5$ from ground-based images
may represent the progenitors of present-day luminous galaxies while forming
their spheroidal stellar component. From the strength of their interstellar UV
absorption lines, and within the assumption that their velocity field is
dominated by gravity, Steidel \etal (1996a) have shown that galaxies at $z\sim
3$ have masses comparable to that of present-day $L_*$ galaxies. Smaller masses
would result if interstellar shocks, local to the star-forming regions,
contribute significantly to the velocity field. While the {\it HST} images of
these galaxies show the presence of compact (half-light radii of 0.2--0.3
arcsec) cores (Giavalisco \etal 1996), from the deeper HDF data is seems that 
the F300W dropouts may be better described as asymmetric or irregular
objects (Abraham \etal 1996; van den Bergh \etal 1996).

The observations described above may confirm the rationale 
that galaxies form stars from the inside out, i.e., that spheroidal systems
have assembled rather early, far beyond the disk epoch at $z\sim 1$. At the
star formation density levels inferred from the HDF images at $\langle
z\rangle=2.75$ (see eq. [\hdfone]) about $15\%$ of the observed mass density of
metals at $z=0$ would have been formed during the ``spheroid epoch'' at $z\gta
2$. This fraction could increase if the star formation is shrouded in dust or
takes place in galaxies small enough individually to fall below the HDF source
detection limit. On the other hand, since the metals we observe being formed
are a substantial fraction of the entire metal content of galaxies, {\it it
appears that star formation regions remain largely unobscured by dust
throughout much of galaxy formation.} 

With the data presented here, two arguments which have implications for models
of galaxy formation can be made more quantitative.

$\bullet$ From Efstathiou \etal (1988), the space density of bright ellipticals
today is $n(>L_*)=2.4\times 10^{-4}\,$Mpc$^{-3}$. If a significant fraction of
their stellar population formed in a single burst of duration 1 Gyr early in the
history of the universe, a comparable number density of objects should be
observed at high-$z$ while forming stars at rates in excess of $50-100\msun$
yr$^{-1}$. We find that none of the HDF F300W dropouts spectroscopically
confirmed by Steidel \etal (1996b) has a SFR in excess of $20\msun$ yr$^{-1}$
(Salpeter IMF). Assuming a probed redshift interval between 2 and 3.5, 
we can set an upper limit to the comoving density of very high star-forming
galaxies at $\langle z\rangle=2.75$, $n(>20\msun {\rm yr}^{-1})<6\times
10^{-5}\,$Mpc$^{-3}$. Hence there appears to be a deficit of very bright
objects relative to the expectations of the standard early-and-rapidly-forming
picture for spheroidal systems. 

$\bullet$ In hierarchical models of galaxy formation, small objects form first
and merge together to make larger ones. While much of the activity associated
with star formation and galaxy merging occurs at relatively low redshifts
(White \& Frenk 1991), there is no period when bulges and ellipticals form
rapidly as single units and are very bright: rather, galaxies become
progressively less luminous, more numerous, and more compact at earlier epochs.
A common prediction of hierarchical models appears to be the steepening with
lookback time of the luminosity function due to the large increase of the
abundance of objects forming stars at modest rates (e.g, Cole \etal 1994). It
is then interesting to compare, as a function of redshift, the ratio of the
space densities of galaxies forming stars at rates in excess of, say, 1 and 10
$\msun$ yr$^{-1}$. From the Gallego \etal (1995) \Ha survey, we estimate
$n(>1\msun {\rm yr}^{-1})/n(>10\msun {\rm yr}^{-1})\approx 50$ at the present
epoch. A similar value is found to characterize, at $\langle z \rangle=2.75$,
the distribution of SFRs of the F300W dropouts in the HDF.
Thus, unless a substantial fraction of the modest star-forming galaxies at
these early times are red either because are relatively old or because are
reddened by dust, there seems to be little evidence in our sample for a large 
enhancement of their space density relative to the abundance of high
star-forming objects.

\subsectionbegin{5.3. Chemical Enrichment of the Intergalactic Medium}

There is one piece of evidence pointing towards a metal ejection
density at $z\gta 3$ which is at least as large as derived in equation
(\hdfone), namely the observations of a significant cosmological mass density
of metals associated with QSO absorption systems at high-$z$. We shall
focus here on the numerous \Lya forest clouds. Recent spectra at
high S/N and resolution obtained with the Keck
telescope (Tytler \etal 1995; Cowie \etal 1995) have shown that 50\%--60\% of
the Ly$\alpha$ clouds with $\log N_{\rm HI}>14.5$ have undergone some chemical
enrichment, as evidenced by weak, but measurable \CIV absorption lines. The
typical inferred metallicities, $Z_{\rm IGM}$, range from 0.003 to 0.01 of
solar values, subject to uncertainties of photoionization models. Because the
\Lya clouds have large filling factors and contain a significant fraction of
the baryons in the universe, the new data are strong evidence for a widespread
distribution of metals in the IGM. These metals may have been produced in
situ or in dense gaseous regions of galaxies; the metal-enriched gas was then
expelled from the regions of star formation to large distances consistent with
the sizes of the \Lya clouds. 

Associated with the latter type of enrichment a characteristic MER density can
be estimated as \eqnam{\digm} 
$$ 
{\dot \rho}_{Z,{\rm IGM}}(3)\approx (4\times 10^{-4}\msunits)
\left({\Omega_{\rm IGM}\over 0.05}\right)\left({Z_{\rm IGM}\over
0.005Z_\odot}\right) \left({f_{\rm inj}\over 0.5}\right)^{-1}
\left({\Delta t\over t_H}\right)^{-1}, \eqno(\new)
$$
where $\Omega_{\rm IGM}$ is the baryonic density parameter of the \Lya cloud
phase, $f_{\rm inj}$ is the fraction of heavy elements injected into the IGM
during a timescale $\Delta t$, and $t_H$ is the Hubble time at $z=3$. If QSO
absorption systems trace the bulk of star formation occurring in galaxies at
high redshifts (Lanzetta, Wolfe, \& Turnshek 1995; Pei \& Fall 1995; Madau \&
Shull 1996), it is therefore plausible that a population of dwarf galaxies,
actively forming stars at $z\gta 3$ at a rate comparable to the limit inferred
from the HDF data, may be responsible for the contamination of the IGM at high
redshifts. Such galaxies have shallow gravitational potential wells, thus
allowing easy ejection of the heavy elements produced during their active
star-forming phase. 

\subsectionbegin{5.4. The Ionizing Background at $z\gta 4$}

In the last few years growing evidence has accumulated on the existence of a
significant decline in the space density of bright QSOs beyond $z\sim 3$
(Schneider, Schmidt, \& Gunn 1994; Warren, Hewett, \& Osmer 1994; Kennefick,
Djorgovski, \& de Carvalho 1995). This shortfall, estimated to be about a
factor of 5, poses serious problems to the idea that a quasar-dominated UV
background is responsible for maintaining the intergalactic gas in a highly
ionized state above $z\sim 4$ (e.g., Haardt \& Madau 1996). It may be of some
interest, in this context, to speculate on the possible contribution of our
population of star-forming ``spheroids'' to the metagalactic flux at early
epochs (see also Songaila \etal 1990). A minimum value for the intensity of 
the ionizing background at $z=4$ is given by \eqnam{\jl} 
$$ 
J_L(4)={hc\over 4\pi}n_{\rm H,crit}(4)\Omega_{\rm IGM}\approx 2.4\times
10^{-22}\left({\Omega_{\rm IGM}\over 0.05}\right) \uvunits, \eqno(\new) 
$$ 
where $n_{\rm H,crit}$ is the closure hydrogen density. 
This equation reflects the underlying physics of the ionization process:
ignoring recombinations, at least one UV photon is required per hydrogen atom
to photoionize the IGM. The ionizing radiation flux is largely local, as
sources at higher redshifts are severely absorbed by intervening clouds. At
$z=4$, the attenuation length is only $\Delta l\approx 13\,$Mpc (Madau 1992).
The required (proper) volume emissivity at 1 ryd is then $4\pi J_L/\Delta l$.
Collecting, and converting as usual an ionizing luminosity density into 
a MER per unit comoving volume, we obtain \eqnam{\dion}
$$
{\dot \rho}_{Z,{\rm ion}}(4)\approx (2\times 10^{-4}\msunits)
\left({\Omega_{\rm IGM}\over 0.05}\right)\left({f_{\rm esc}\over
0.5}\right)^{-1}, \eqno(\new) 
$$
where now $f_{\rm esc}$ is the escape fraction into the IGM of Lyman-continuum
photons. A comparison between equations (\hdftwo) and (\dion) provides some
rationale for the hypothesis that a population of galaxies at $z\sim 4$,
producing metals at a rate which is a few times higher than the lower limit
derived from the HDF data, might rival quasars as a source of photoionization
of the IGM. Larger star formation rates would be required if a significant
fraction of the UV radiation emitted from stars cannot escape into the
intergalactic space, $f_{\rm esc}\ll 1 $.

\vskip 0.5truecm
\bigskip
We have benefited from discussions with M. Fall, D. Hogg, C. Leitherer, S.
Lilly, Y. Pei, and M. Shull. Support for this work was provided by NASA through
grant AR-06337.10-94A from the Space Telescope Science Institute, which is
operated by the Association of Universities for Research in Astronomy, Inc.,
under NASA contract NAS5-26555. M.G. acknowledges support from the Hubble
Fellowship program through grant number HF-01071.01-94A. C.C.S. acknowledges
support from the Sloan Foundation and from the NSF through grant AST-9457446. 

\page
%table 1
%\page
%table 2 
%\page
%table 3 
%\page

\centerline{\bf REFERENCES}

\refindent{Abraham, R.~G., Tanvir, N.~R., Santiago, B.~X., Ellis, R.~S., 
Glazebrook, K., \& van den Bergh, S. 1996, MNRAS, in press}

\apjj{Babul, A., \& Ferguson, H.~C. 1996}{458}{100}
 
\refindent{Bertelli, G., Bressan, A., Chiosi, C., Fagotto, F., \& Nasi, E.
1994, A\&AS, 106, 275} 

\refindent{Bressan, A., Chiosi, C., \& Fagotto, F. 1994, ApJS, 94, 63}

\apjj{Bruzual, A.~G., \& Charlot, S. 1993}{405}{538}

\apjj{Calzetti, D., Kinney, A.~L., \& Storchi-Bergmann, T. 1994}{429}{582}

\apjj{Charlot, S., Worthey, G., \& Bressan, A. 1996}{457}{625}

\mnrass{Clegg, R.~E.~S., \& Middlemass, D. 1987}{228}{759}

\refindent{Cohen, J.~G., Cowie, L.~L., Hogg, D.~W., Songaila, A., Blandford, 
R., Hu, E.~M., \& Snopbell, P. 1996, submitted to ApJ}

\mnrass{Cole, S., Arag\'on-Salamanca, A., Frenk, C.~S., Navarro, J.~F., \&
Zepf, S.~E. 1994}{271}{781} 

\refindent{Cowie, L.~L. 1988, in The Post-Recombination Universe, ed. N. Kaiser
and A. Lasenby (NATO Advanced Science Institute Series), p. 1}

\refindent{Cowie, L. L., Songaila, A., Kim, T.-S., \& Hu, E. M. 1995, AJ, 
109, 1522}

\apjj{da Costa, L.~N., \etal 1994}{424}{L1}

%\refindent{Dickinson, M.~E., \etal 1996, in preparation}

\refindent{Djorgovski, S., \& Thompson, D. 1993, in IAU Symp. 149,
The Stellar Populations in Galaxies, eds. A. Renzini \& B. Barbuy 
(Dordrecht: Kluwer), 337}

\mnrass{Efstathiou, G., Ellis, R.~S., \& Peterson, B.~A. 1988}{232}{431}

\refindent{Ellis, R.~S., Colless, M., Broadhurst, T., Heyl, J., \& Glazebrook,
K. 1996, MNRAS, 280, 235}

\refindent{Ferguson, H.~C., \etal 1996a, in preparation}

\refindent{--------- . 1996b, in preparation}

\apjj{Ferguson, H.~C., \& McGaugh, S.~S 1995}{440}{470}

\ajj{Gallagher, J.~S., Bushouse, H., \& Hunter, D.~A. 1989}{97}{700}

\apjj{Gallego, J., Zamorano, J., Arag{\'o}n-Salamanca, A., \& Rego, M. 1995}
{455}{L1}

\refindent{Giavalisco, M., Steidel, C.~C., \& Macchetto, F.~D 1996, ApJ, 
in press}

%\refindent{Giavalisco, M., \etal 1996b, in preparation}

\mnrass{Glazebrook, K., Ellis, R.~S., Santiago, B.~X., \& Griffith, R. 1995}
{275}{L19}

\apjj{Guhathakurta, P., Tyson, J.~A., \& Majewski, S.~R. 1990}{357}{L9}

\refindent{Gwyn, S.~D.~J., \& Hartwick, F.~D.~A. 1996, ApJ, in press}

\apjj{Haardt, F., \& Madau, P. 1996}{461}{20}

\ajj{Kennefick, J.~D., Djorgovski, S.~G., \& de Carvalho, R.~R. 1995}{110}{6}

\apjj{Kennicutt, R.~C. 1992}{388}{310}

\refindent{Kinney, A.~L., Bohlin, R.~C., Calzetti, D., Panagia, N., 
\& Wyse, R.~F.~G. 1993, ApJS, 86, 5}

\apjj{Kruk, J.~W., \etal 1995}{454}{L1}

\mnrass{Kuijken, K., \& Gilmore, G. 1989}{239}{605}

\refindent{Kurucz, R.~L. 1992, private communication}

\refindent{Lanzetta, K.~M., Yahil, A., \& Fern{\' a}ndez-Soto, A. 1996,
Nature, 381, 759}

\apjj{Lanzetta, K.~M., Wolfe, A.~M., \& Turnshek, D.~A. 1995}{440}{435}

\apjj{Leitherer, C., Ferguson, H.~C., Heckman, T.~M., \& Lowenthal, 
J.~D. 1995}{454}{L19}

\apjj{Lilly, S.~J., Tresse, L., Hammer, F., Crampton, D., \& Le F{\'e}vre,
O. 1995}{455}{108}

\apjj{Lilly, S.~J., Le F{\'e}vre, O., Hammer, F., \& Crampton, D.,
1996}{460}{L1}

\apjj{Loveday, J., Peterson, B.~A., Efstathiou, G., Maddox, S.~J. 1992}{390}
{338}

\apjj{Madau, P. 1992}{389}{L1}

\apjj{--------- . 1995}{441}{18}

\apjj{Madau, P., \& Shull, J.~M. 1996}{457}{551}

\aaa{Maeder, A. 1992}{264}{105}

\apjj{Marzke, R.~O., Huchra, J.~P., \& Geller, M.~J. 1994}{428}{43}

\aaa{M{\o}ller, P., \& Jakobsen, P. 1990}{228}{299} 

\refindent{Moustakas, L., Zepf, S., \& Davis, M. 1996,
http://astro.berkeley.edu/davisgrp/HDF} 

\refindent{Oke, J.~B. 1974, ApJS, 27, 21}

\refindent{Ortolani, S., Renzini, A., Gilmozzi, R., Marconi, G., Barbuy, 
B., Bica, E., \& Rich, M.~R. 1995, Nature, 377, 701}

\refindent{Osterbrock, D.~E. 1989, Astrophysics of Gaseous Nebulae and 
Active Galactic Nuclei (Mill Valley: University Science Books)} 

\refindent{Pagel, B.~E.~J., \& Edmunds, M.~G. 1981, ARA\&A, 19, 77}

\apjj{Pei, Y.~C. 1992}{395}{130}

\apjj{Pei, Y.~C., \& Fall, S.~M. 1995}{454}{69}

\apjj{Press, W.~H., Rybicki, G.~B., \& Schneider, D.~P. 1993}{414}{64}

\apjj{Pritchet, C.~J., \& Hartwick, F.~D.~A. 1990}{355}{L11}

\apjj{Rao, S. \& Briggs, F. 1993}{419}{515}

\apjj{Renzini, A., Ciotti, L., D' Ercole, A., \& Pellegrini, S. 1993}
{419}{52}

\aaa{Renzini, A., \& Voli, M. 1981}{94}{175}

\apjj{Salpeter, E.~E. 1955}{121}{161}

\apjj{Schade, D., Lilly, S.~J., Crampton, D., Le F{\'e}vre,
O., Hammer, F., \& Tresse, L. 1995}{451}{L1}

\ajj{Schneider, D.~P., Schmidt, M., \& Gunn, J.~E. 1991}{101}{2004}

\ajj{--------- . 1994}{107}{1245}

\apjj{Songaila, A., Cowie, L.~L., \& Lilly, S.~J. 1990}{348}{371}

\apjj{Steidel, C.~C., Giavalisco, M., Pettini, M., Dickinson, M., 
\& Adelberger, K.~L. 1996a}{462}{L17}

\refindent{Steidel, C.~C., Giavalisco, M., Dickinson, M., \& Adelberger, 
K.~L. 1996b, AJ, in press}

\ajj{Steidel, C.~C., \& Hamilton, D. 1992}{104}{941}
 
\ajj{--------- . 1993}{105}{2017}

\ajj{Steidel, C.~C., Pettini, M., \& Hamilton, D. 1995}{110}{2519}

\apjj{Storrie-Lombardi, L.~J., McMahon, R.~G., Irwin, M.~J., \& Hazard, C.
1994}{427}{L13}

\refindent{Sutherland, R.~S., \& Shull, J.~M. 1996, in preparation}

\aaa{Thielemann, F.~K., Nomoto, K., \& Yokoi, K. 1986}{158}{17}

\refindent{Tinsley, B.~M. 1980, Fund. Cosmic Phys., 5, 287}

\refindent{Tytler, D., Fan, X.-M., Burles, S., Cottrell, L., Davis, C.,
Kirkman, D., \& Zuo, L.  1995, in QSO Absorption Lines, Proc. ESO Workshop,
ed. G. Meylan (Heidelberg: Springer), 289}

\refindent{van den Bergh, S., Abraham, R.~G., Ellis, R.~S., Tanvir, N.~R.,
Santiago, B.~X., \& Glazebrook, K. 1996, AJ, in press} 

\mnrass{van der Marel, R.~P. 1991}{253}{710}

\apjj{Warren, S.~J., Hewett, P.~C., \& Osmer, P.~S. 1994}{421}{412}

\apjj{White, S.~D.~M., \& Frenk, C.~S. 1991}{379}{25}

\refindent{Williams, R.~E., \etal 1996, AJ, in press}

\refindent{Woosley, S.~E., \& Weaver, T.~A. 1995, ApJS, 101, 181}

\apjj{Yoshii, Y., \& Peterson, B. 1994}{436}{551}

\vfill\eject

\centerline{\bf FIGURE CAPTIONS}
\bigskip
\ni {\bf Figure 1:}
({\it a}) Mean cosmic transmission for a source at $z_{em}=2.5, 3.5$, and $4.5$
({\it solid lines}), as a function of observed wavelength. The characteristic 
staircase profile is due to continuum blanketing from the Lyman series. 
Also plotted are the response functions of the four broad passbands, 
F300W ({\it dotted line}), F450W ({\it short-dashed line}), F606W
({\it long-dashed line}), and F814W ({\it dash-dotted line}) used for the 
{\it Hubble Deep Field}. ({\it b}) Magnitude increments $\Delta U_{300}$
({\it dotted line}), $\Delta B_{450}$ ({\it short-dashed lines}), $\Delta 
V_{606}$ ({\it long-dashed line}), and $\Delta I_{814}$ ({\it dash-dotted
line}), derived by integrating the mean cosmic transmission over the
corresponding bandpass, as a function of the emission redshift. 

\ni {\bf Figure 2:}
({\it a}) Apparent synthetic colors of a star-forming galaxy with constant 
SFR (age 0.3~Gyr) plotted as a function of the emission redshift. Each pair 
of curves depicts, from top to bottom, the attenuated colors and the colors
observed in the case of negligible intergalactic absorption. {\it Solid lines:}
$U_{300}-B_{450}$. {\it Dashed lines:} $B_{450}-V_{606}$. {\it Dotted lines:}
$V_{606}-I_{814}$. ({\it b}) The colors of the four brightest starburst
galaxies observed by {\it HUT} (see text for details). {\it Empty dots:}
$U_{300}-B_{450}$. {\it Filled dots:} $B_{450}-V_{606}$. {\it Triangles:}
$V_{606}-I_{814}$. All spectra have been reddened by  intergalactic absorption. 

\ni {\bf Figure 3:} $\ub$ vs. $\bv$ for model galaxies.  A total
of 103879 synthetic spectra of galaxies representing a wide range
of ages, star formation histories, metallicities, dust contents, 
and redshifts, were folded through the HDF bandpasses. The galaxies
shown as small points are at redshifts less than 2 or redshifts
greater than 3.5. Galaxies shown as large solid circles are 
those in the redshift range $2 < z < 3.5$ with ages less than
$10^8$ yr and extinctions $A_B < 1$. Large x's are galaxies
in the same redshift range that have ages greater than $10^8$ yr
or $A_B > 1$. Our color selection criteria are shown as the large polygon.
Galaxies within the polygon in the HDF observations are selected as likely
candidates for $2 < z < 3.5$ objects. The selection criteria are $\ub > 1.3$,
$\ub > \bi + 1.2$, and $\bi < 1.5$.

\ni {\bf Figure 4:}  Selection efficiency for F300W dropouts.
This figure provides a measure of color selection in isolating galaxies with
$2< z < 3.5$. The solid histogram shows the fraction of models in each redshift
interval ($\Delta z = 0.2$) that meet the color selection criteria. The dotted
histogram shows the fraction of models with ages less than $10^8$ yr and
extinctions $A_B < 2$ that meet the selection criteria. The color selection
criteria are extremely efficient at identifying relatively unobscured
star-forming galaxies with $2< z < 3.5$, and separating them from the chaff
at other redshifts. 

\ni {\bf Figure 5:} $\bv$ vs. $\vi$ for model galaxies. The galaxies
shown as small points are at redshifts less than 3.5 or redshifts
greater than 4.5. Galaxies shown as large solid circles are those in the
redshift range $3.5 < z < 4.5$ with ages less than $10^8$ yr and extinctions
$A_B < 1$. Large x's are galaxies in the same redshift range that have ages
greater than $10^8$ yr or $A_B > 1$. Our color selection criteria are shown as
the large polygon. Galaxies within the polygon in the HDF observations are
selected as likely candidates for $3.5 < z < 4.5$ objects. The selection
criteria are $\bv>1.7(\vi)+0.7$, $\bv > 1.5$, $\bv < 3.5(\vi)+1.5$, and $\vi<
1.5$. 

\ni {\bf Figure 6:} Selection efficiency for F450W dropouts. This figure
provides a measure of color selection in isolating galaxies with $3.5 < z <
4.5$. The solid histogram shows the fraction of models in each redshift interval
($\Delta z = 0.2$) that meet the color selection criteria. The dotted histogram
shows the fraction of models with ages less than $10^8$ yr and extinctions $A_B
< 2$ that meet the selection criteria. The primary source of contamination for
the sample is galaxies at slightly lower redshifts with intrinsically red
colors. The dashed line shows the fraction of the models with ages greater than
$10^9$ yr or $A_B > 2$ that meet the selection critera. If such red objects at
$2 < z < 3.5$ are very common, then our derived metal formation rate at higher
redshifts may be an overestimate. However, we suspect that such old or dusty
interlopers are much less common than blue star-forming galaxies at slightly
higher redshift. 

\ni {\bf Figure 7:} $U_{300}-B_{450}$ vs. $B_{450}-I_{814}$ color-color plot
of galaxies in the {\it Hubble Deep Field} satistying magnitude ranges given
in the text. ({\it a}) Objects undetected in F300W (with signal-to-noise $<$1)
are plotted as triangles at the $1\sigma$ lower limits to their 
$U_{300}-B_{450}$ colors. Symbol size scales with the $I_{814}$ 
magnitude of the object. The dashed lines outline the selection region within 
which we identify candidate $2<z<3.5$ objects. ({\it b}) Location in the
color-color diagram of the more than 60 galaxies in the HDF which have known
Keck/LRIS spectroscopic redshifts from Steidel \etal (1996b), Cohen \etal 
(1996), and Moustakas \etal (1996). 

\ni {\bf Figure 8:} $B_{450}-V_{606}$ vs. $V_{606}-I_{814}$ color-color plot
of galaxies in the {\it Hubble Deep Field} satistying magnitude ranges given
in the text. ({\it a}) The symbol types and sizes are as described for Fig. 
7{\it a}. The selection area we define for candidate $3.5<z<4.5$ galaxies is
outlined with dashed lines. ({\it b}) Same as Fig. 7{\it b}. 

\ni {\bf Figure 9:} Element and star formation history of the universe. The 
data points from various surveys provide a measurement or a lower limit to the
universal metal ejection density, $\dot \rho_Z$, as a function of redshift. 
For a Salpeter IMF, to translate $\dot \rho_Z$ into a total star formation
density, $\dot \rho_*$, a factor of 42 should be applied. {\it Triangle:}
Gallego \etal (1995). {\it Filled dots:} Lilly \etal (1996). {\it Diagonal 
cross:} lower limit from Steidel \etal (1996a). {\it Filled squares:} lower
limits from the {\it Hubble Deep Field} images. The dashed line depicts the
fiducial rate, ${\dot \Sigma_Z}$, given by the mass density of metals observed
today divided by the present age of the universe (see text for details).  A
flat cosmology with $q_0=0.5$ and $H_0=50\kmsmpc$ has been assumed. 

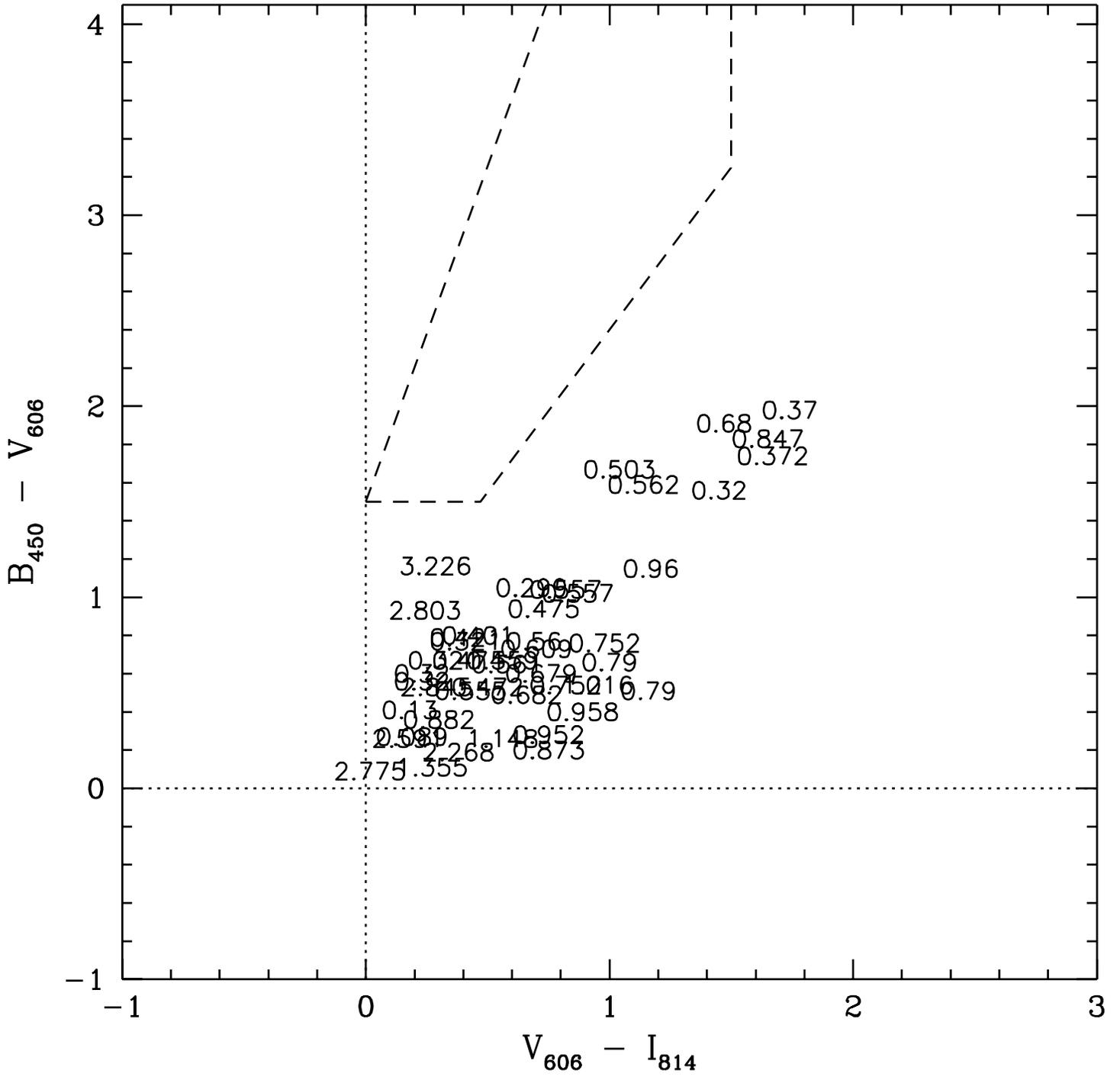
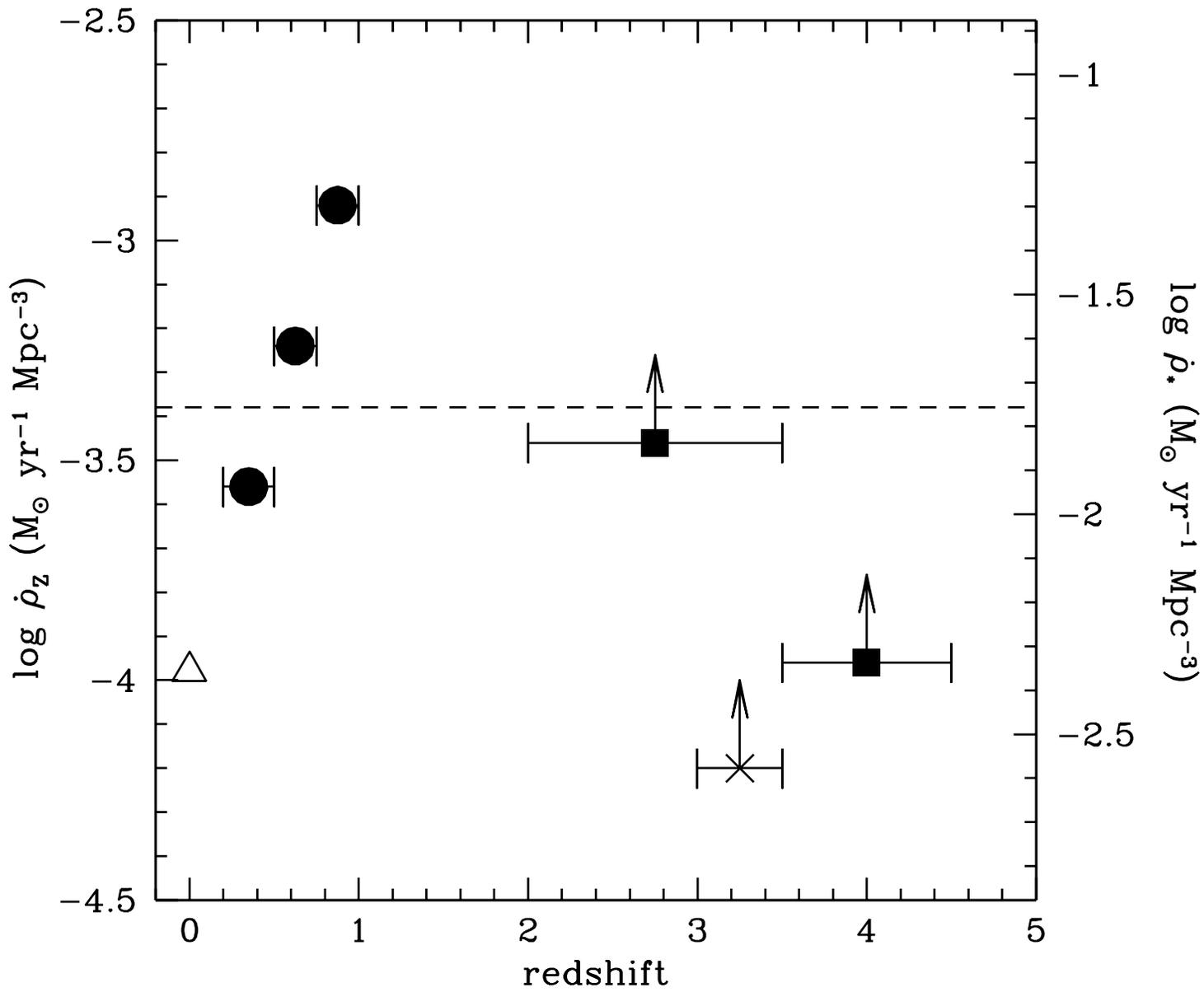
\bye